\documentclass[runningheads,fleqn]{svmult}
\usepackage[dvips]{hyperref}
\usepackage{makeidx}   
\usepackage{graphicx}  
\usepackage{subeqnar}  
\usepackage{multicol}  
\usepackage{taphys}    
\begin{document}
\title*{{\sc ac}-Calorimetry at High Pressure and Low Temperature}
\toctitle{{\sc ac}-Calorimetry at High Pressure \protect\newline
and Low Temperatures}
%
%
\titlerunning{{\sc ac}-Calorimetry at High Pressure and Low Temperature}
%
\author{Heribert Wilhelm}

\authorrunning{H. Wilhelm}
%
%
\institute{Max-Planck-Institut f\"ur Chemische Physik fester Stoffe,
N\"othnitzer Str. 40, 01187 Dresden}

\maketitle              

%
%

\begin{abstract}
Recent developments of the \textsc{ac}-calorimetric technique adapted
for the needs of high pressure experiments are discussed. A
semi-quantitative measurement of the specific heat with a
Bridgman-type of pressure cell as well as a diamond anvil cell is
possible in the temperature range 0.1~K$<T<10$~K. The pressure
transmitting medium used to ensure good pressure conditions determines
to a great extent via its thermal conductivity the operating frequency
and thus the accessible temperature range. Investigations with
different pressure transmitting media for $T>1.5$~K reveal for solid
He a cut-off frequency which is considerably higher than for
steatite. Experiments below 1~K and pressures above 10~GPa clearly
show that the pressure dependence of the linear temperature
coefficient of the specific heat can be measured. It is in qualitative
agreement to a related quantity obtained quasi-simultaneously by
electrical resistivity measurements on the same sample.
\end{abstract}

%
%

The specific heat ($C$) is an important thermodynamic quantity. Its
temperature dependence can deliver hints about microscopic energy
scales and provides a powerful tool to identify phase transitions. In
this respect temperature ($T)$ dependent measurements are an
indispensable means not only for experimentalists. This has triggered
the development of different and very sophisticated technical
realizations to obtain $C(T)$ from the millikelvin range up to very
high temperature. The available methods can be divided in two
categories. Adiabatic techniques are considered as the most accurate
way to estimate the absolute value of $C(T)$. They require sample
masses of several grams and the subtraction of the addenda, i. e., the
specific heat of sample holder and thermometer. Among the
non-adiabatic (or dynamic) methods, {\sc ac}-calorimetry is a suitable
technique for samples with masses well below one milligram. The
specific heat can be measured with very high sensitivity, despite the
small masses. However, the absolute accuracy which can be achieved is
less than for the adiabatic methods.

Adiabatic techniques are used to detect pressure-induced phase
transitions or to investigate the evolution of electronic properties
as the unit cell volume is reduced. The sample masses needed demand
large volume pressure cells, such as a piston-cylinder cell.  With
this technique the accessible pressure range is, however, limited to
about 3.5~GPa. Very often it would be desirable for the pressure range
to be extended. In this case an anvil-type of pressure cell is the
only alternative.  Such a high pressure tool demands a much smaller
sample volume which makes an adiabatic measurement a hopeless
venture. Thus, {\sc ac}-calorimetry is an ideal method to be used for
pressures beyond the limit of piston-cylinder cells. 
%
%

\section{ {\sc ac}-calorimetry adapted for high pressure}
\label{sec:experimental}

\begin{figure}
\center{\includegraphics[width=110mm,clip]{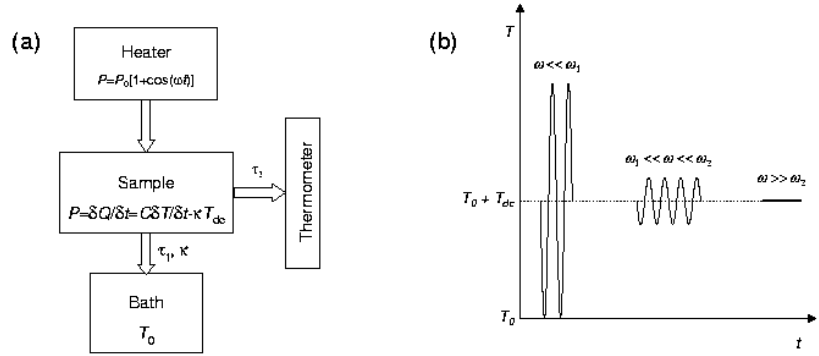}}
\caption{(a) Sketch of a general {\sc ac}-calorimetric assembly. The
sample, thermal bath, thermometer, and heater are in contact by a
thermal link with thermal conductivity $\kappa$. $\tau_1$ is a measure
of the thermal relaxation between sample and bath; $\tau_2$ comprises
the relaxation of thermometer, heater, and sample. (b) Sample
temperature $T(t)$ for different frequency domains: For $\omega \ll
\omega_1$,$T_{ac}\equiv T_{dc}\propto P_0/\kappa$ is not frequency
dependent and is a measure of the thermal conductivity $\kappa$. In
the range $\omega_1\ll \omega \ll \omega_2$, the amplitude of the
\textsc{ac}-part $T_{ac}\propto (\omega C)^{-1}$ depends on the
measuring frequency and yields the specific heat of the sample. At
$\omega\gg \omega_2$, $T_{ac}$ is strongly reduced. Independent of the
frequency, the mean sample temperature is $T=T_0+T_{dc}$.}
\label{fig:schema}
\end{figure}

The general set-up of the {\sc ac}-calorimetric technique for measuring the
specific heat is sketched as a simplified model in Fig.~\ref{fig:schema}(a).
The sample is thermally excited by an oscillating heating power $P=P_0[1+
\cos(\omega t)]$, e.g. generated by a current of frequency $\omega/2$ through
a resistance heater. The temperature oscillations at frequency $\omega$ are
detected with a thermometer attached to the sample. Sullivan and Seidel
\cite{Sullivan68} obtained a relation among the amplitude $T_{ac}$ of the
temperature oscillations and the specific heat $C$ of the sample:
\begin{equation}
T_{ac} = \frac{P_0}{\omega C}\left\{1+ \frac{1}{\omega^2\tau_1^2}
                               + \omega^2\tau_2^2\right\}^{-1/2}~.
\label{eq:tac}
\end{equation}

\noindent This equation contains the time constants $\tau_1=C/\kappa$
and $\tau_2$, with $\kappa$ the thermal conductivity of the thermal
link between sample and temperature bath (see
Fig.~\ref{fig:schema}(a)). It was derived in the ideal case, when the
heat capacity of thermometer, heater, and heat link between sample and
temperature bath are negligible and assuming a perfect coupling
between heater, sample, and thermometer. The measured value of
$T_{ac}$ depends on the measuring frequency $\omega$
(Fig.~\ref{fig:schema}(b)): At low frequency
($\omega\ll\omega_1=\kappa/C$) the mean sample temperature
is above the bath temperature $T_0$ by $T_{ dc}\propto
P_0/\kappa$. The recorded temperature oscillation $T_{ac}$ yields the
specific heat of the sample if the frequencies are in the range
$\omega_1\ll\omega\ll\omega_2 = 1/\tau_2$. The possibility of
tuning both the amplitude and the frequency of the excitation is the
main advantage of this method; as long as $\kappa$ can be made small
enough, the sensitivity of the measurement does not depend on the mass
of the sample.

%
%

This technique was employed by several groups
\cite{Bonilla74,Baloga77,Eichler79,Eichler80} to investigate the
pressure dependence of the specific heat. The conditions for the {\sc
ac}-technique in a pressure cell are far away from being ideal. In
particular the thermal properties of the pressure transmitting medium
have to be taken into account. This was done by Baloga and
Garland~\cite{Baloga77} for the case of high gas densities and low
sample thermal conductivities. In their accessible temperature range
(245~K~$<T<~$300~K) the general relation between $T_{ac}$ and $C$ for
the \textsc{ac}-calorimetric expression (\ref{eq:tac}) can be
recovered if the product of specific heat and thermal conductivity of
the pressure transmitting medium is negligible with respect to that of
the sample.  Then the heat wave does not propagate too far into the
pressure transmitting medium and its specific heat does not contribute
too much to $T_{ac}$. Typical frequencies are of the order of
1~Hz. Eichler and Gey \cite{Eichler79} were the first to use the
\textsc{ac}-technique for metallic samples in a piston-cylinder cell
($p_{max}\approx 3.5$~GPa) at low temperature (1.3~K$<T<7$~K). Here,
the sample was embedded in diamond powder. It acts as pressure
transmitting medium and provided the thermal resistance between the
sample and the pressure cell. The measuring frequency was 120~Hz.

Pressures well above 3.5~GPa can only be achieved with opposed anvils,
i.e., with a clamped Bridgman anvil technique or a diamond anvil cell
(DAC). In Bridgman cells, the anvils are often made out of tungsten
carbide (WC) or synthetic diamond and the pressure chamber consists of
pyrophyllite (a sheet silicate, Al$_2$Si$_4$O$_{10}$(OH)$_2$). The
sample is in between two disks of e.g., the soft mineral steatite
(3MgO$\cdot$4SiO$_2\cdot$H$_2$O) which acts as pressure transmitting
medium. In a DAC a metallic gasket contains the sample and the
pressure transmitting medium. Compared to a Bridgman cell a DAC
comprises several advantages. First of all the pressure range can be
extended easily to 50~GPa. Furthermore, the transparent anvils give
optical access to the sample and the pressure can be determined with
the ruby fluorescence method. Finally, the most important point is the
possibility using He as pressure transmitting medium. With respect to
hydrostatic pressure conditions, solidified He is an ideal medium
since it is highly plastic and inert. However, these desirable
features might mislead in underestimating the efforts in the elaborate
assembly of the {\sc ac}-calorimetric circuit in a DAC.

%
%
\begin{figure}
\center{\includegraphics[width=100mm,clip]{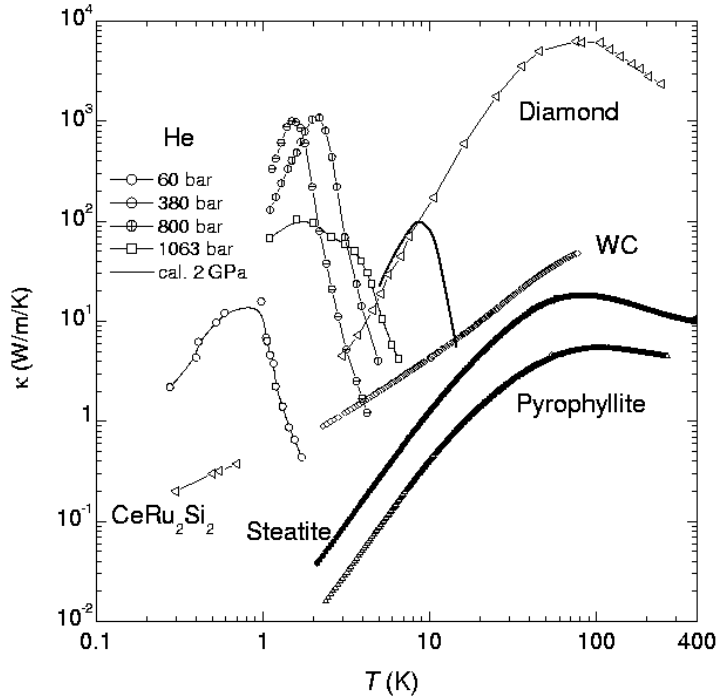}}
\caption{Thermal conductivities $\kappa(T)$ of different materials
used in high pressure devices with opposed anvils. WC and diamond are
often used as anvils. Data for diamond (type Ib) are taken from
\cite{Landolt}. Pyrophyllite and steatite serve as gasket and pressure
transmitting medium, respectively. Due to its plasticity solid He
permits homogeneous pressure conditions. The $\kappa(T)$ data for He
are taken from \cite{Webb52} (60~bar) and \cite{Seward69} (380~bar and
800~bar). Unpurified He (1063~bar \cite{Seward69}) has a
significantly different $\kappa(T)$. At the indicated pressures the
crystals were grown. The solid line represents a calculated
$\kappa(T)$ of He at 2~GPa (see text). $\kappa(T)$ of CeRu$_2$Si$_2$
\cite{Amato} stands in for the thermal conductivity of heavy Fermion
compounds at low temperature.}
\label{fig:kappaall}
\end{figure}

The feasibility of the {\sc ac}-technique at pressures well above the
limit of the piston-cylinder cells has to be tested, regardless of the
type of high pressure cell. From the general principle of the {\sc
ac}-calorimetry (Fig.~\ref{fig:schema}) it is evident that the main
challenge are the unknown thermal properties of the pressure
transmitting medium. To shed some light on this, the thermal
conductivity of the two preferred media, steatite and He will be
discussed qualitatively in the following.

A priori it is not evident if the pressure media could satisfy the
assumed requirements in the deduction of (\ref{eq:tac}) because little
is known about their thermal conductivity under pressure. To get an
overview of $\kappa(T)$ of the materials used in a Bridgman device,
the thermal conductivity of WC, pyrophylitte, and steatite at ambient
pressure have been measured (Fig.~\ref{fig:kappaall}). Literature data
of diamond \cite{Landolt} and solid He at different pressure
\cite{Webb52,Seward69} are also depicted in
Fig.~\ref{fig:kappaall}. At low temperatures $\kappa(T)$ of steatite
can be a factor 10$^4$ smaller than that of solid He at about 0.1~GPa.
Moreover, the purity of He significantly affects the shape and size of
$\kappa(T)$. For very pure He \cite{purity} the maximum thermal
conductivity can be one order of magnitude higher than for unpurified
He at almost the same pressure \cite{Seward69}. Fortunately, it is
very likely that the solidified He in the pressure chamber of a DAC is
polycrystalline and contains impurities, brought in during the filling
procedure. Therefore, $\kappa(T)$ might not reach the high values
expected for pure He at the same pressure. As was pointed out in
\cite{Seward69} the maximum value of $\kappa(T)$ occurs roughly at
$\Theta_D/50$ ($\Theta_D$: Debye temperature). Since He is highly
compressible, $\Theta_D$ and thus, the maximum of $\kappa(T)$, rapidly
increases with pressure. As a result the value of $\kappa(T)$ at,
e.g. 1~K, could decrease considerably. In order to get a rough
estimate of $\kappa(T)$ of He at several GPa, the following
assumptions are made: (i) The low temperature slope remains unchanged
at high pressure. (ii) The maximum value of $\kappa(T)$ at $T_{max}$
does not increase with pressure. (iii) $T_{max}\approx\Theta_D/50$ can
be estimated using $\Theta_D(V)=\Theta_D(V_0)\left
[\frac{V}{V_0}\right]^{\gamma}$ with $\gamma=2.4$
\cite{Dugdale64}. (iv) The maximum value of $\kappa(T)$ for He in a
DAC might be comparable to that of impure He at about 0.1~GPa
(Fig.~\ref{fig:kappaall}). Based on the equation of state for solid He
\cite{Mills80} the density for low temperature and 2~GPa is inferred
to be about $V\approx 6$~cm$^3$/mole. This density together with
$\Theta_D(V_0)\approx 90$~K at $V_0=11.77$~cm$^3$/mole
\cite{Dugdale64} yields $T_{max}\approx 9$~K. Then $\kappa(T)$ at
2~GPa (line in Fig.~\ref{fig:kappaall}) can be estimated with the
assumptions specified above.

$\kappa(T)$ of steatite and He shown in Fig.~\ref{fig:kappaall}
clarifies the differences in an {\sc ac}-calorimetric experiment with
these pressure media. The cut-off frequency in the case of steatite
changes continuously since $\kappa(T)$ is a monotonic function below
10~K. Moreover, it is very likely that its shape will not be effected
strongly by pressure. Thus, $\omega_1$ at a given temperature should
slightly vary with pressure. For steatite $\kappa(T)=a T^{2.3}$, with
$a=6.7\times 10^{-3}$~W/m/K$^{3.3}$, is a good approximation of the
data below 8~K. Together with heat capacity of a typical heavy Fermion
compound like CeRu$_2$Ge$_2$ \cite{Wilhelm99} or CePd$_2$Ge$_2$
\cite{Wilhelm02} a cut-off frequency $\omega_1/(2\pi)\approx 100$~Hz
at ambient pressure is calculated. $\kappa(T)$ of pressurized He,
however, varies drastically with temperature and pressure. Comparing
$\kappa(T)$ of He at 2~GPa and 4.2~K with that of steatite at ambient
pressure shows that $\omega_1$ will be roughly a factor 10 larger and
of the order of several kHz. At these frequencies severe constraints
are put on the homogeneity of the temperature in the sample. A
homogeneous temperature distribution in the sample is given if the
thermal wavelength $\lambda_{th}\propto \sqrt{\kappa/(C\omega)}$ is of
the order of the sample thickness (typically about 30~$\mu$m). This
condition is already fulfilled at about $\omega/(2\pi)\approx 1$~kHz
for the compounds mentioned above. Nevertheless, these frequencies are
well below $\omega_2$ since metallic samples and thermometer ensure
high thermal conductivity.

These considerations show that an {\sc ac}-calorimetric measurement of
metallic samples enclosed in solid He is more difficult for
1~K$<T<10$~K as for the same sample embedded in steatite. Outside this
temperature interval $\kappa(T)$ of He can be as small as that of
steatite at ambient pressure and it might expected that $T_{ac}$ is
dominated by the specific heat of the sample. If the assumptions made
above hold for even higher pressure, the cut-off frequency could be
reduced significantly and the \textsc{ac}-calorimetry in a DAC would become
feasible in a larger temperature range.

%
%

\section{{\sc ac}-calorimetry in different pressure environments}
\label{sec:pressuremedia}

The previous section illuminated the general aspects of the
\textsc{ac}-calorimetry and contemplated the frequency domain in which
experiments could be conducted. Two independent experiments
\cite{Bouquet00,Salce00,Demuer00} using the same compound but in
different pressure devices and pressure transmitting media provide
experimental information about the cut-off frequency. Both
investigations explored pressures up to 8~GPa and temperatures in the
range 1.5~K$<T<10$~K.

\begin{figure}
\center{\includegraphics[width=90mm,clip]{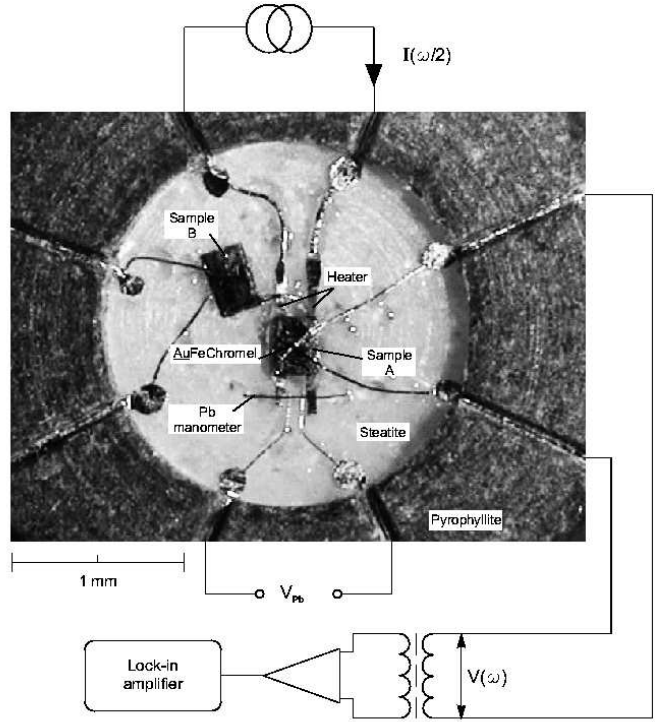}}
\caption{Top view of the inner part of a Bridgman-type of pressure
cell before closing. Two samples of CeRu$_2$Ge$_2$ are arranged for an
\textsc{ac}-calorimetric experiment. Sample A is placed on top of the
heater wires but is insulated from them. Sample B is in contact with a
metallic foil and thermally linked to the heater through a
Au-wire. The Chromel-\underline{Au}Fe thermocouples measure the sample
temperature. The Pb-wire serves as pressure gauge. The entire assembly
is mounted on a disk of steatite.}
\label{fig:cellcrg}
\end{figure}

Figure~\ref{fig:cellcrg} shows the pressure chamber of the Bridgman
cell before closing the device. The typical thickness of the sample,
thermocouple and heating wires are 20, 12, and 3~$\mu$m,
respectively. Two different ways of supplying the heat to the samples
were tested. For sample~A a thin electrical insulation (4--5~$\mu$m of
an epoxy/Al$_2$O$_3$ mixture) prevents electrical contact with the
heater but still established a good thermal contact. Sample B is set
apart on a metallic (Pb) foil, electrically (and thus thermally)
linked to the heater through a gold wire. No heating current passes
through this sample. In the course of the experiment it turned out that
the configuration A provided a homogeneous temperature distribution
whereas the configuration B was ensuring hydrostatic pressure conditions.

%
%

The heating power was chosen in such a way that the temperature
oscillations were in the range 2~mK$<T_{ac}<20$~mK. They were measured
with a \underline{Au}Fe/Au thermocouple (Au + 0.07~at\% Fe). The
thermovoltage $V_{ ac}$ arises from the temperature difference between
the sample (at $T_0 + \Delta T$) and the edge of the sample chamber
(at $T_0$) \cite{Jaccard98}. The thermovoltage was amplified at room
temperature in two stages and read by means of a lock-in detection
referred to the frequency of the heating current. However, two
potential drawbacks should not be concealed: (i) the temperature of
the samples is measured with a thermocouple, under the assumption that
the ambient pressure calibration holds at high pressure. (ii) The
total amount of heat supplied to the samples is not known, despite the
resistive heating. This prevents so far the acquisition of absolute
values for the specific heat.

%
%
\begin{figure}
\includegraphics[width=57mm,clip]{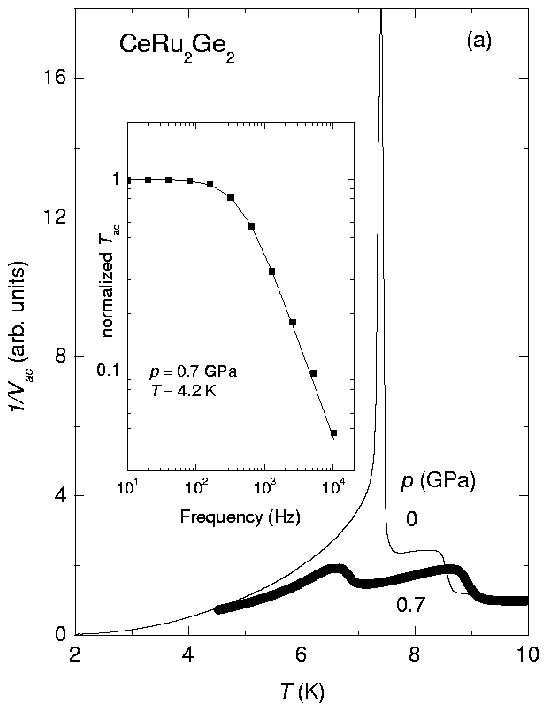}
\includegraphics[width=57mm,clip]{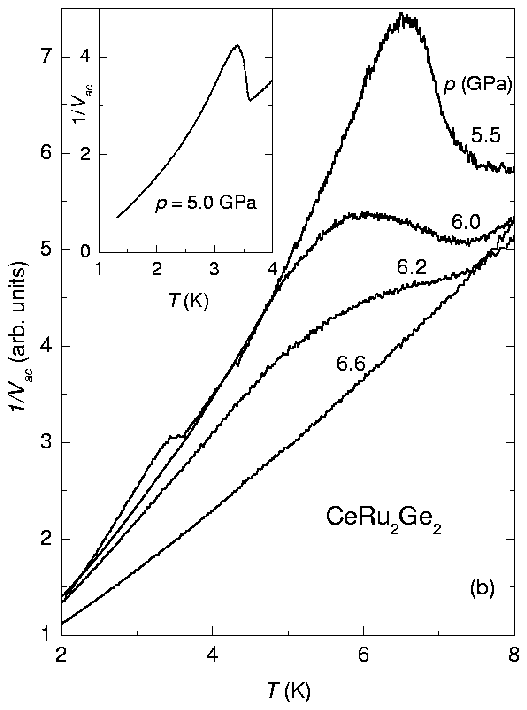}
\caption{(a) Comparison of the inverse of the lock-in signal,
$1/V_{ac}\propto C$, of CeRu$_2$Ge$_2$ enclosed in steatite at 0.7~GPa
and the specific heat $C/T$ measured with a relaxation method at
ambient pressure \cite{Wilhelm99}. The data sets are normalized at
10~K. A frequency test at 4.2~K is depicted in the inset
($\omega_1\approx 450$~Hz). (b) Temperature dependence of
$1/V_{ac}\propto C$ of CeRu$_2$Ge$_2$ above 5~GPa. The pronounced
feature related to the antiferromagnetic transition is suppressed by
pressure. The inset shows the specific heat at 5.0~GPa with an anomaly
at low temperature. The feature can still be seen near 3.5~K at
5.5~GPa (main figure).}
\label{fig:cp_crg}
\end{figure}

CeRu$_2$Ge$_2$ exhibits two magnetic phase transitions at ambient
pressure leading to large features in the specific heat. Together with
the well known influence of pressure on these transitions
\cite{Wilhelm99} this compound is a good candidate for testing
\textsc{ac}-calorimetry at high pressure. Figure~\ref{fig:cp_crg}(a)
shows the result of the \textsc{ac}-measurements at 0.7~GPa in
comparison to the specific heat obtained by a relaxation method at
ambient pressure. Pressure slightly shifted the transition
temperatures as expected from the ($T,p$) phase diagram
\cite{Wilhelm99}. The height of the specific heat jump at the second
order transition ($T_{ N} \approx$~9~K) represents 47\% of the total
signal compared to 51\% for the ambient pressure curve. This indicates
that $T_{ac}$ is dominated by the heat capacity of the sample. An
additional support for this statement is given by a frequency
test. According to (\ref{eq:tac}) the relation $T_{ac}\propto
1/\omega$ for $\omega\gg\omega_1$ should hold, which is indeed
observed (inset Fig.~\ref{fig:cp_crg}(a)). A fit of a low pass filter
to the data yields $\omega_1/(2\pi)=450$~Hz. Frequency tests at
various temperatures and pressures are a necessary task to determine
$\omega_1$ and to ensure the validity of the relation between $T_{ac}$
and the specific heat of the sample. The height of the first order
transition ($T_{ C} \approx$~7~K) is very sensitive to any
distribution of $T_{ C}$ and should not compared to the peak in
$C_p(T)$ at ambient pressure. Moreover, {\sc ac}-calorimetry is not
the proper tool to measure a latent heat \cite{Wen92}
since it only detects the reversible part at frequency $\omega$ on a
temperature scale $T_{ac}$. Nevertheless, the position of a first
order transition can be detected by an {\sc ac}-calorimetric
measurement.

The {\sc ac}-calorimetry data of CeRu$_2$Ge$_2$ above 5~GPa shown in
Fig.~\ref{fig:cp_crg}(b) demonstrate the potential of this method. The
influence of pressure on the antiferromagnetic transition is visible and the
deduced $T_N(p)$ data agree with the ($T,p$) phase diagram extracted from
transport measurements \cite{Wilhelm99}. A critical pressure $p_c\approx
7$~GPa is necessary to suppress the long-range magnetic order. The broadening
of the antiferromagnetic transition is very likely related to intrinsic
effects although a small pressure inhomogeneity could be partly responsible
for it. In addition to this transition an anomaly at lower temperature was
resolved (inset of Fig.~\ref{fig:cp_crg}(b)). These measurements were the
first to show that this anomaly seen so far only by transport measurements
\cite{Wilhelm99}, has thermodynamic origin and is a bulk property.

Working with a DAC allowed Demuer and coworkers \cite{Demuer00} to use a
different way to supply the oscillating heat power to the sample. They
attached an optical fiber to the DAC and heated the sample with the light of
an Ar-ion laser. It was chopped mechanically at frequencies up to 3~kHz. The
temperature oscillations of a Au-Chromel thermocouple bonded directly on the
sample by spark welding were measured with a lock-in amplifier. In this
experiment CeRu$_2$Ge$_2$ was enclosed with solidified He. The cut-off frequency
was estimated to 4~kHz at 0.5~GPa and 7~K. This value supports the estimated
order of magnitude for $\omega_1$ in the case of pressurized He given in
Sec.~\ref{sec:experimental}. The high thermal conductivity of He limits the
application of the {\sc ac}-method at low pressures. Nevertheless, the
magnetic phase transitions could be observed although only a part of the
signal at the fixed measuring frequency of 1.5~kHz was due to the specific
heat of the sample. In this investigation an increased width of the transition
was also established. Intrinsic effects seem to be responsible for this if the
good hydrostatic pressure conditions in the experiment are kept in mind. In
addition, a similar broadening in specific heat experiments at ambient and low
pressure have been reported \cite{Fisher91,Bogenberger95} when $T_N$ is pushed
to zero temperature either by doping or pressure.

The analysis of the thermal conductivity data in
Sec.~\ref{sec:experimental} suggests that He could be used as a
pressure transmitting medium at low temperature even at pressures of a
few GPa. This presumption is corroborated by the results of Holmes and
coworkers \cite{Holmes03}. With a combined measurement of electrical
resistivity and {\sc ac}-calorimetry the heavy Fermion superconductor
CeCu$_2$Si$_2$ was investigated down to 0.1~K for pressures up to
7~GPa. The jump in the {\sc ac}-signal caused by the entrance into the
superconducting state provided a semi-quantitative measure of the
sample specific heat. The onset of the specific heat occurred when the
resistive transition was completed and affirms the bulk property of
the superconducting state.

%
%

\section{{\sc ac}-calorimetry below 1~K and beyond 10~GPa}
\label{sec:cepdge}

\begin{figure}
\center{\includegraphics[width=115mm,clip]{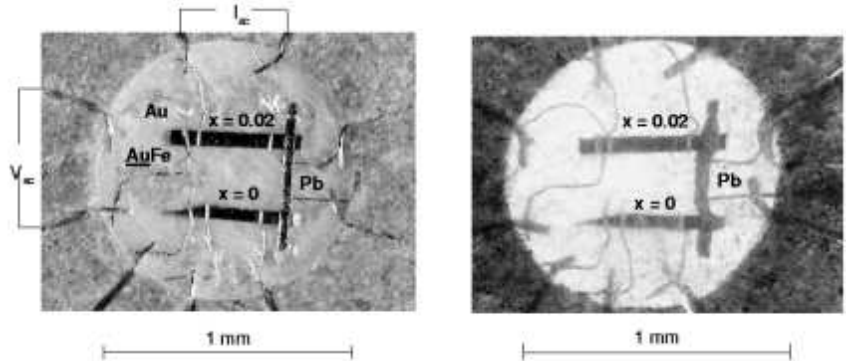}}
\caption{Left: Pressure cell before closing with two samples of
CePd$_{2+x}$Ge$_{2-x}$ ($x=0$ and 0.02) and the pressure gauge
(Pb-foil).  The \textsc{ac}-calorimetric circuit is mounted on one
sample ($x=0.02$). The temperature oscillations are read with an
additional \underline{Au}Fe wire when an \textsc{ac}-heating current
is applied. Right: The pressure cell after pressure release from 22~GPa. The
almost unchanged configuration shows the reliability of the pressure
device.}
\label{fig:cpgcell}
\end{figure}

A demonstration of the feasibility of the \textsc{ac}-technique below
1~K and pressures well above 10~GPa is the experiment on
CePd$_{2.02}$Ge$_{1.98}$ in a Bridgman-type of pressure cell
\cite{Wilhelm02}. The set-up of the experiment was chosen in such a
way that electrical resistivity and \textsc{ac}-calorimetry could be
performed on the same crystal. This makes it possible to check whether
an anomaly in $T_{ ac}$ is related to the sample or not with an
independent electrical resistivity
measurement. Figure~\ref{fig:cpgcell} shows the arrangement in the
pressure chamber. It contains two different samples of the
solid-solution CePd$_{2+x}$Ge$_{2-x}$, but only one of them ($x=0.02$)
was connected for the \textsc{ac}-experiment. The sample was heated
with a current supplied through Au-wires attached to the sample. This
reduces the components in the pressure chamber and avoids a pressure
gradient due to the heat wires. With this arrangement it is also
possible to calibrate the \underline{Au}Fe/Au thermocouple up to very
high pressure and over a wide temperature range \cite{Wilhelm02}. It
was observed that the absolute thermopower $S(T)$ of \underline{Au}Fe
at 4.2~K and 1.0~K at 12~GPa is about 20\% smaller than the values at
ambient pressure. These rather small changes show that the results are
not affected qualitatively if the ambient pressure values of $S(T)$
are used. Thus, the drawback of a missing temperature calibration for
the thermocouple mentioned in Sec.~\ref{sec:pressuremedia} could be in
principle eliminated.
%
%
\noindent
The reliability of the pressure cell is obvious as can be
seen in Fig.~\ref{fig:cpgcell}. After the pressure was released
from 22~GPa the overall shape of the pressure cell as well as its
initial diameter were almost unchanged and the distance between the
voltage leads increased by less than 5\%.

%
%
\begin{figure}
\center{\includegraphics[width=100mm,clip]{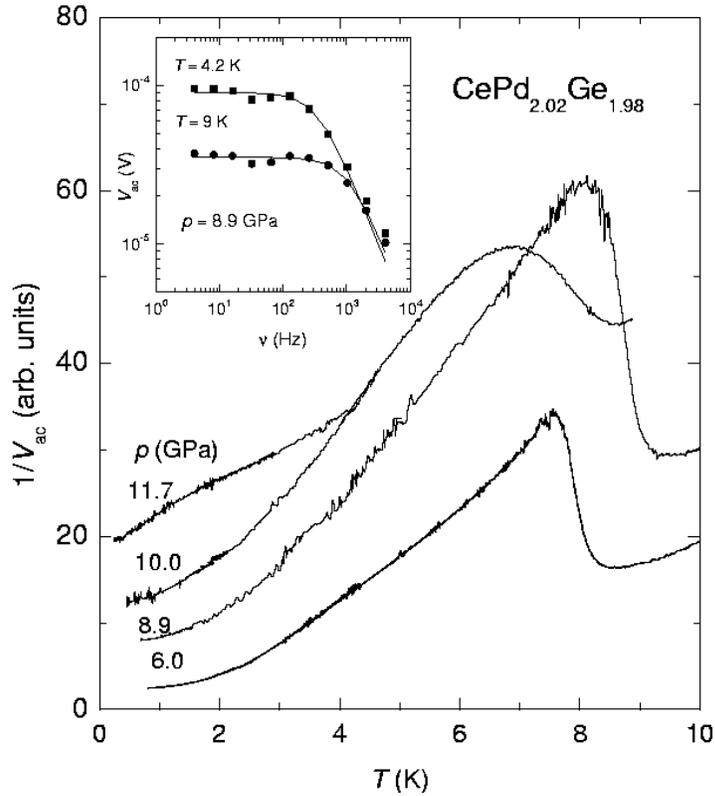}}
\caption{Temperature dependence of the inverse lock-in voltage $V_{
ac}$ of CePd$_{2.02}$Ge$_{1.98}$. The entrance into the
antiferromagnetically ordered state is clearly visible. Inset:
Frequency test at $p=8.9$~GPa for different temperatures. The solid
lines represent a fit of a low pass filter to the data with cut-off
frequencies $\omega_1/(2\pi)=350$~Hz and 1060~Hz for 4.2~K and 9~K,
respectively.} \label{fig:cp_cpg}
\end{figure}

CePd$_{2.02}$Ge$_{1.98}$ was chosen because in its stoichiometric form
it is the Ge-doped counterpart of the antiferromagnetically ordered
heavy Fermion compound CePd$_2$Si$_2$ ($T_N=10$~K). The latter system
enters a superconducting ground state when the magnetic order is
suppressed ($p_c=2.7$~GPa) \cite{Grosche96}. Applying pressure to
CePd$_{2.02}$Ge$_{1.98}$ ($T_N=5.16$~K \cite{Wilhelm02}) should
increase $T_N$ to a maximum and then it should approach zero
temperature. The aim of the {\sc ac}-calorimetric measurement was to
extract the electronic contribution to the specific
heat. Figure~\ref{fig:cp_cpg} shows the inverse of the registered
lock-in signal $V_{ ac}$ below 10 K at various pressures.  The
pronounced anomaly in $1/V_{ ac}(T)$ for pressures between 6.0 GPa and
10 GPa is caused by the entrance into the antiferromagnetically
ordered phase.  The height of the anomaly decreases and it becomes a
very broad feature as the system approaches $p_c=11.0$~GPa
\cite{Wilhelm02}. A similar broadening upon approaching the critical
pressure was reported for CePd$_2$Si$_2$, despite the lower pressure
and the use of He as pressure transmitting medium
\cite{Demuer02}. Recalling the increased transition width reported in
Sec.~\ref{sec:pressuremedia} it is very likely that this is an
intrinsic phenomenon. Two frequency tests at $p=8.9$~GPa are shown in
the inset of Fig.~\ref{fig:cp_cpg}. A fit of a low pass filter to the
data yield cut-off frequencies of $\omega_1/(2\pi)=350$~Hz and 1060~Hz
for $T=4.2$~K and 9~K, respectively. Assuming the validity of
$\omega_1=\kappa/C$, these values and the $1/V_{ac}$ data at the
corresponding temperatures result in
$\kappa(4.2$~K)/$\kappa(9$~K)$\approx 0.2$, almost the same ratio as
at ambient pressure (see Fig.~\ref{fig:kappaall}). Hence, pressures up
to 9~GPa seem to have a weak effect on $\kappa(T)$ of steatite below
10~K.

%
%
\begin{figure}
\center{\includegraphics[width=115mm,clip]{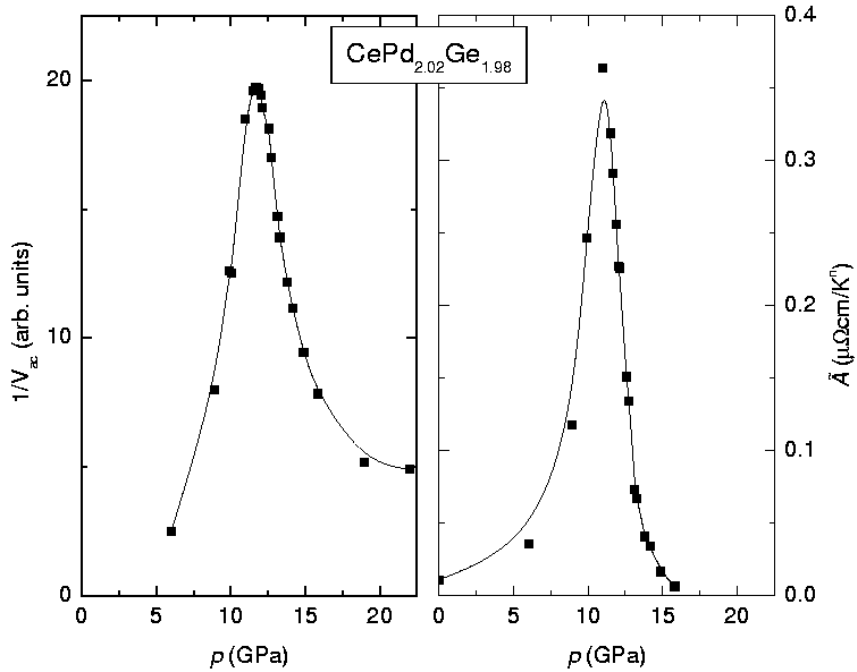}}
\caption{Left: Pressure dependence of the inverse lock-in voltage $V_{
ac}$ of CePd$_{2.02}$Ge$_{1.98}$ obtained at the lowest temperatures
reached in each run. The maximum is attained at a pressure very close
to the critical pressure where the magnetic ordering temperature is
pushed to zero. Right: The temperature coefficient $\tilde{A}$ of the
electrical resistivity. It shows an anomaly at the
magnetic/non-magnetic phase transition.} \label{fig:gamma}
\end{figure}

The most important observation in this experiment is the pressure
dependence of the value of $1/V_{ ac}$ taken at about 0.3~K. The
inverse of the lock-in voltage, $1/V_{ac}$, strongly increases,
reaches a maximum in the vicinity of $p_c$ and levels off at high
pressure (Fig.~\ref{fig:gamma}). The critical pressure was inferred
from the $\tilde{A}(p)$-anomaly in the temperature dependence of the
electrical resistivity $\rho(T)=\rho_0 + \tilde{A}T^n$, with $\rho_0$
the residual resistivity, and the fitting parameters $\tilde{A}$ and
$n$ \cite{Wilhelm02}. Below 1~K, $1/V_{ac}$ is proportional to $C/T$,
since the temperature dependence of the absolute thermopower,
$S(T)\propto T$, is a fairly good assumption. Above this temperature
the $S(T)$ dependence is certainly different and $1/V_{ ac}$ has to be
interpreted with caution. Thus, $1/V_{ ac}(T)$ at low temperature can
be regarded as a direct measure of the electronic correlations. The
pronounced pressure dependence of $1/V_{ ac}$ shows that the
electronic correlations are considerably enhanced as pressure
approaches $p_c$ and that the signal originates mainly from the
sample. However, above 15~GPa, the pressure dependence is not strong
enough to follow the $\tilde{A}(p)$-dependence according to the
empirical Kadowaki-Woods relation \cite{Kadowaki86}. A possible reason
for this deviation might be that at these pressures $V_{ac}$ does not
represent entirely the heat capacity of the sample. A step towards a
quantitative measure of the specific heat at these conditions would be
to achieve a control of the supplied heating power and the thermal
contact between sample and pressure transmitting medium. Nevertheless,
the strong pressure dependence of $1/V_{ ac}$ at low temperature is
reminiscent to $\tilde{A}(p)$ and is a motivation for further studies.

\section{Conclusions}
The {\sc ac}-calorimetric technique adapted for high pressure
experiments at low temperature ($T<10$~K) was discussed. The
oscillating sample temperature provides the specific heat of the
sample if the measuring frequency is above the cut-off frequency
$\omega_1=\kappa/C$. It is determined by the thermal conductivity of
the pressure transmitting medium and the specific heat of the
sample. A qualitative estimate of $\kappa(T)$ for steatite and solid
He, the two preferred pressure media was made. The cut-off frequency
for steatite is less than 1~kHz whereas several kHz was inferred for
solid He (at $\approx 2$~GPa and 4.2~K). An experimental confirmation
of the order of magnitude for these values was found for pressures up
to 7~GPa and temperatures in the range of 1.5~K$<T<10$~K. The large
values in the case of He put a temperature limit for the use of a DAC
whereas a Bridgman-type of high pressure cell can be used below 10~K
and pressures well above 10~GPa. Due to the strong pressure dependence
of $\kappa(T)$ of He, the maximum in $\kappa(T)$ shifts towards higher
temperature. This might open the low temperature region for the {\sc
ac}-calorimetric method also for a DAC. These promising results build
up the hope of a quantitative understanding of the {\sc
ac}-calorimetry and interesting specific heat data under extreme
conditions might be expected.\\

\noindent {\bf Acknowledgments}\\

\noindent The work presented here is the result of a collaboration with F.
Bouquet, A. Demuer, A. Holmes, D. Jaccard, A. Junod, and Y. Wang. I am
grateful to them for many fruitful and stimulating discussions. The assistance
of A. Bentjen in the thermal conductivity measurements at the MPI CPfS is
acknowledged.

%


\begin{thebibliography}{8.}
\addcontentsline{toc}{section}{References}

\bibitem{Sullivan68}P. F. Sullivan and G. Seidel, Phys. Rev. {\bf 173}, 679 (1968).

\bibitem{Bonilla74}A. Bonilla and C. W. Garland, J. Phys. Chem. Solids, {\bf 35}, 871 (1974).

\bibitem{Baloga77}J. D. Baloga and C. W. Garland, Rev. Sci. Instrum. {\bf 48}, 105 (1977).

\bibitem{Eichler79}A. Eichler,  and W. Gey, Rev. Sci. Instrum. {\bf 50}, 1445 (1979).

\bibitem{Eichler80}A. Eichler, H. Bohn, and W. Gey, Z. Phys. B {\bf 38}, 21 (1980).

\bibitem{Landolt}Landolt B\"ornstein Vol. III/17a: Physics of Group
IV Elements and III-V Compounds, ed.: O. Madelung, Springer, Berlin,
p. 357 (1982).

\bibitem{Webb52}F. J. Webb, K. R. Wilkinson, and J. Wilks, Proc. Roy. Soc. A, \textbf{214}, 546 (1952).

\bibitem{Seward69}W. D. Seward, D. Lazarus, and S. C. Fain, Jr.,
Phys. Rev. {\bf 178}, 345 (1969).

\bibitem{Amato}A. Amato, Ph.D Thesis, University of Geneva, 1988.

\bibitem{purity}According to \cite{Seward69} purified (unpurified) He refers to a
chemical impurity level less than 15 ppm with an (no) additional purification
by an adsorption trap cooled to 63~K.

\bibitem{Dugdale64}J. S. Dugdale and J. P. Franck, Phil. Trans.
Roy. Soc. London A, {\bf 257}, 1 (1964).

\bibitem{Mills80}R. L. Mills, D. H. Liebenberg, and J. C. Bronson, Phys. Rev.
B \textbf{21}, 5137 (1980).

\bibitem{Wilhelm99}H. Wilhelm, K. Alami-Yadri, B. Revaz, and
 D. Jaccard, Phys. Rev. B {\bf 59}, 3651  (1999).

\bibitem{Wilhelm02}H. Wilhelm and D. Jaccard, Phys. Rev. B {\bf 66},
064428 (2002).

\bibitem{Bouquet00}F. Bouquet, Y. Wang, H. Wilhelm, D. Jaccard, and A. Junod, Solid State Commun.
{\bf 113}, 367 (2000).

\bibitem{Salce00}B. Salce, J. Thomasson, A. Demuer, J.J. Blanchard,
J.M. Martinod, L. Devoille, and A. Guillaume, Rev. Sci. Insturm. {\bf 71},
2461 (2000).

\bibitem{Demuer00}A. Demuer, C. Marcenat, J. Thomasson, R. Calemczuk, B. Salce,
P. Lejay, D. Braithwaite, and J. Flouquet, J. Low Temp. Phys. {\bf 120}, 245
(2000).

\bibitem{Jaccard98}D. Jaccard, E. Vargoz, K. Alami-Yadri, and
H. Wilhelm, Rev. High Pressure Sci. Technol. {\bf 7}, 412 (1998).

\bibitem{Wen92} X. Wen, C. W. Garland, R. Shashidhar, P. Barois,
Phys. Rev. B {\bf 45}, 5131 (1992).

\bibitem{Fisher91}R. A. Fisher, C. Marcenat, N. E. Phillips, P. Haen,
F. Lapierre, P. Lejay, J. Flouquet, and J. Voiron, J. Low Temp. Phys. {\bf 84},
49 (1991).

\bibitem{Bogenberger95}B. Bogenberger and H. v. L\"ohneysen,
Phys. Rev. Lett. {\bf 74}, 1016 (1995).

\bibitem{Holmes03}A. T. Holmes, A. Demuer, and D. Jaccard, Acta Phys. Pol. B {\bf 34}, 567 (2003).

\bibitem{Grosche96}F. M. Grosche, S. R. Julian, N. D. Mathur, and
G. G. Lonzarich, Physica B {\bf 223\&224}, 50 (1996).

\bibitem{Demuer02}A. Demuer, A. T. Holmes, and D. Jaccard, J. Phys.:
Condens. Matter {\bf 14}, L529 (2002).

\bibitem{Kadowaki86}K. Kadowaki and S. B. Woods, Solid State Commun. {\bf 58}, 507 (1986).






\end{thebibliography}
\end{document}